\documentclass[article,twocolumn,preprintnumbers,superscriptaddress,nofootinbib]{revtex4-1}

\usepackage{enumerate}
\usepackage{mathrsfs,amsmath,amssymb}
\usepackage{natbib}
\usepackage{graphicx}
\usepackage{cancel}
\usepackage[normalem]{ulem}
\usepackage[pdftex,bookmarks,linktocpage,pdfpagelabels,plainpages=false,hyperfigures,linkcolor=blue,citecolor=blue]{hyperref} 
\hypersetup{colorlinks=true}

\newcommand\be{\begin{equation}}
\newcommand\ee{\end{equation}}
\newcommand\bea{\begin{eqnarray}}
\newcommand\eea{\end{eqnarray}}

\newcommand\kev{\rm keV}
\newcommand\gev{\rm GeV}
\newcommand\m{\rm m}
\newcommand\s{\rm s}

\begin{document}
\bibliographystyle{apsrev4-1}

\hfill \preprint{ MI-TH-2017}

\title{Explaining the XENON1T excess with Luminous Dark Matter  
}

\author{Nicole F.~Bell}
\email{n.bell@unimelb.edu.au}
\affiliation{ARC Centre of Excellence for Dark Matter Particle Physics, School of Physics, The University of Melbourne, Victoria 3010, Australia}

\author{James B.~Dent} 
\email{jbdent@shsu.edu}
\affiliation{Department of Physics, Sam Houston State University, Huntsville, TX 77341, USA}

\author{Bhaskar Dutta}
\email{dutta@physics.tamu.edu}
\affiliation{Mitchell Institute for Fundamental Physics and Astronomy,
Department of Physics and Astronomy, Texas A\&M University, College Station, TX 77843, USA}
   
\author{Sumit Ghosh}
\email{ghosh@tamu.edu}
\affiliation{Mitchell Institute for Fundamental Physics and Astronomy, 
Department of Physics and Astronomy, Texas A\&M University, College Station, TX 77843, USA}

\author{Jason Kumar}
\email{jkumar@hawaii.edu}
\affiliation{Department of Physics$,$~ University~ of~ Hawaii$,$~ Honolulu$,$~ Hawaii~ 96822$,$~ USA}

\author{Jayden L.~Newstead}
\email{jayden.newstead@unimelb.edu.au}
\affiliation{ARC Centre of Excellence for Dark Matter Particle Physics, School of Physics, The University of Melbourne, Victoria 3010, Australia}

\begin{abstract}

We show that the excess in electron recoil events seen by 
the XENON1T experiment can be explained by relatively 
low-mass Luminous Dark Matter candidate.  The dark matter 
scatters inelastically in the detector (or the surrounding 
rock), to produce a heavier dark state with a 
$\sim 2-3~\kev$ mass splitting.  This heavier state 
then decays within the detector, producing a peak in 
the electron recoil spectrum which is a good fit to the 
observed excess. We comment on the ability of future 
direct detection experiments to differentiate this model 
from other Beyond the Standard Model scenarios, and from 
possible tritium backgrounds, including the use of diurnal modulation, multi-channel signals etc.,~as possible distinguishing features of this scenario.

\end{abstract}

\maketitle

\section{Introduction}

\par Recently the XENON Collaboration announced an excess of low energy electron recoil events above their expected background \cite{Aprile:2020tmw}. Though this excess may originate from a tritium $\beta$-decay that was previously not included in their background model, the collaboration also examined Beyond the Standard Model (BSM) possibilities including solar axions or a neutrino magnetic moment ($\mu_\nu$) \cite{Bell:2006wi,Bell:2005kz,Borexino:2017fbd}. With a trace amount of tritium ($6.2\pm2.0\times10^{-20}$ mol/mol) added to the background model, the anomaly is explained at 3.2$\sigma$ significance, while the background plus solar axion (background plus $\mu_\nu$) solution provides a 3.5$\sigma$ (3.2$\sigma$) significance fit to the excess within certain parameter ranges. These BSM possibilities lose substantial statistical significance when combined with a tritium component in the fit - down to 2.1$\sigma$ ($0.9\sigma$) for the solar axion ($\mu_\nu$) case. It should also be noted that the axion explanation of the excess is in tension with astrophysical constraints~\cite{DiLuzio:2020jjp}. Additionally, the collaboration examined the possibility of bosonic dark matter, but found no global significance above $3\sigma$.  Other studies of BSM explanations for the 
excess include~\cite{Takahashi:2020bpq, Kannike:2020agf, 1802131, 1802133, 1802130, 1802141,Smirnov:2020zwf,Harigaya:2020ckz,Du:2020ybt,Choi:2020udy,Chen:2020gcl,AristizabalSierra:2020edu,Paz:2020pbc,Buch:2020mrg}.

\par The XENON1T excess is characterized by a peak at $\sim 3~\kev$.  
In this work we consider the possibility that the XENON1T 
excess is generated by the interactions of Luminous Dark Matter (LDM)~\cite{Feldstein:2010su,Pospelov:2013nea,Eby:2019mgs}, with 
a mass splitting in the $\delta \sim 
3~\kev$ range.  
The basic idea is that dark matter scattering is purely 
inelastic, with the dark matter ($\chi$) scattering off 
nuclei 
(either in the detector or in the surrounding overburden) to 
produce an excited dark state ($\chi'$).  The dark state 
then decays ($\chi' \rightarrow 
\chi \gamma$) by the emission of a monoenergetic photon with 
energy $\sim \delta$.  Given the energy resolution of 
XENON1T, the resulting electron recoil spectrum contains a 
peak which is a good fit to the XENON1T excess.

\par The paper is organized as follows.  In Section~\ref{sec:LDM} we briefly review the setup of 
Luminous Dark Matter, and its application to the 
XENON1T excess.  In Section~\ref{sec:results}, we present 
our results.   
In Section~\ref{sec:prospects}, we discuss the prospects 
for future experiments to probe this model.
We conclude with a discussion of our 
results in Section~\ref{sec:summary}.
\\
\section{Luminous Dark Matter} \label{sec:LDM}

Our basic model is a species of Luminous Dark Matter. This is a two-state inelastic dark matter scenario in which the heavier dark state produces photons via its decays. 
Specifically, the cosmological cold dark matter is a particle 
$\chi$ with mass $m_\chi$, and there exists a slightly heavier 
dark state $\chi'$, whose mass exceeds $m_\chi$ by the 
mass splitting $\delta = m_{\chi'} - m_\chi \ll m_\chi$.
The dominant decay of $\chi'$ is through $\chi' \rightarrow 
\chi \gamma$.  Indeed, if $\delta$ is sufficiently small 
and if $\chi'$ and $\chi$ have the same spin, this is the 
only visible decay which will be accessible (a two neutrino final state would also be possible).  Note that, 
if $\delta \ll m_\chi$, then in the rest frame of the $\chi'$ 
we will find $E_\gamma =\delta 
+ {\cal O}(\delta^2 / m_\chi)$.  
 Note that even if 
$\chi'$ decays to $\chi$ and multiple photons, the sum of 
photon energies will be $\delta 
+ {\cal O}(\delta^2 / m_\chi)$, because the outgoing 
$\chi$ will have negligible kinetic energy for 
$\delta / m_\chi \ll 1$. This scenario can emerge if the dark matter is coupled to a mediator, $\phi$, through a $\chi \chi'\phi$ interaction with $\phi$ decaying to $\gamma\gamma$.

In this scenario, dark matter scattering is entirely 
inelastic ($\chi A \rightarrow \chi' A$).  This type of 
purely inelastic scattering arises generically in a variety 
of contexts~\cite{TuckerSmith:2001hy, TuckerSmith:2004jv, Finkbeiner:2007kk, Arina:2007tm, Chang:2008gd, Cui:2009xq, Fox:2010bu, Lin:2010sb, DeSimone:2010tf, An:2011uq, Pospelov:2013nea,Finkbeiner:2014sja, Dienes:2014via, Barello:2014uda, Bramante:2016rdh,Bell:2018pkk, Jordan:2018gcd}.  
For example, 
inelastic scattering mediated by a dark photon with a 
vector coupling to the dark matter is generic 
in any model where dark matter is only charged under 
spontaneously broken continuous symmetries.  The reason 
is that a gauge boson can only couple to a complex degree 
of freedom.  But if all of the continuous symmetries under 
which the dark matter is charged are spontaneously broken, 
then the dark matter is generically expected to split into 
two real degrees of freedom.  Since one cannot form a 
vector current with a single real degree of freedom, the 
dark photon must instead couple to an off-diagonal vector 
current, yielding inelastic scattering.  Moreover, a small mass splitting can be technically natural, e.g., in models where the two dark states form a pseudo-Dirac fermion.

As with the ambient dark matter particles, $\chi$, the $\chi'$ produced from inelastic scattering is non-relativistic. Therefore, the eventual decay of the $\chi'$ yield nearly monoenergetic photons in the frame of the Earth.  This spectrum will have a peak at $\delta$, and a width of roughly $\beta \delta$, where $\beta \sim {\cal O}(10^{-3})$ is the approximate velocity of $\chi'$ in the frame of the Earth.  For our purposes, this is essentially a line signal.  But this monoenergetic signal will be smeared by the energy resolution of the detector.  Note also that, in order for inelastic scattering to be kinematically allowed, one must have $\delta \lesssim m_\chi v^2$; if $\delta \sim {\cal O}(\kev)$, then we must have $m_\chi \gtrsim {\cal O}(\gev)$.

Note that if the lifetime of $\chi'$ is short, ${\cal O}(1 \mu\s)$, it will decay within the XENON1T fiducial volume if the initial inelastic scatter itself took place within this volume.  In that case it is possible for the initial scatter to also produce a detectable signal, either in the form of a nuclear recoil, or an electron recoil via the Migdal effect~\cite{Migdal:1941, Ibe:2017yqa, Dolan:2017xbu, Bell:2019egg}. However, we shall see that for the inelastic cross section required to explain the XENON1T electron recoil excess, the corresponding nuclear recoil signal is below current experimental sensitivity when $m_\chi \alt 15~\gev$.  Additionally, the upscatter could cause some events to be removed due to the multi-scatter veto, reducing the detection efficiency. Lastly, the Migdal process will be a subleading effect, since only a very small fraction of inelastic scatters will produce a Migdal electron, whereas every inelastic scatter will produce a photon via $\chi'$ decay.  

If the lifetime of the $\chi'$ is longer, then it is not 
necessary for the initial scatter to even occur within 
the detector.  Instead the dark matter could scatter within 
the surrounding rock, with the produced $\chi'$ decaying 
within the detector.  Provided that the decay length of 
the $\chi'$ is at most comparable to the length of the 
overburden, one would find that the rate of $\chi'$ decay 
in the detector is similar to the rate of dark matter 
scattering in the detector\footnote{In this case, there 
will be some differences between the rate of scattering 
and the rate of $\chi'$ decay in the the detector, due to 
the differing densities and compositions of the detector 
versus the surrounding material.  But this has little 
effect on our main result.}.


\section{The Xenon excess and LDM } \label{sec:results}

\par
\begin{figure}[tbp]
\includegraphics[width=8.cm]{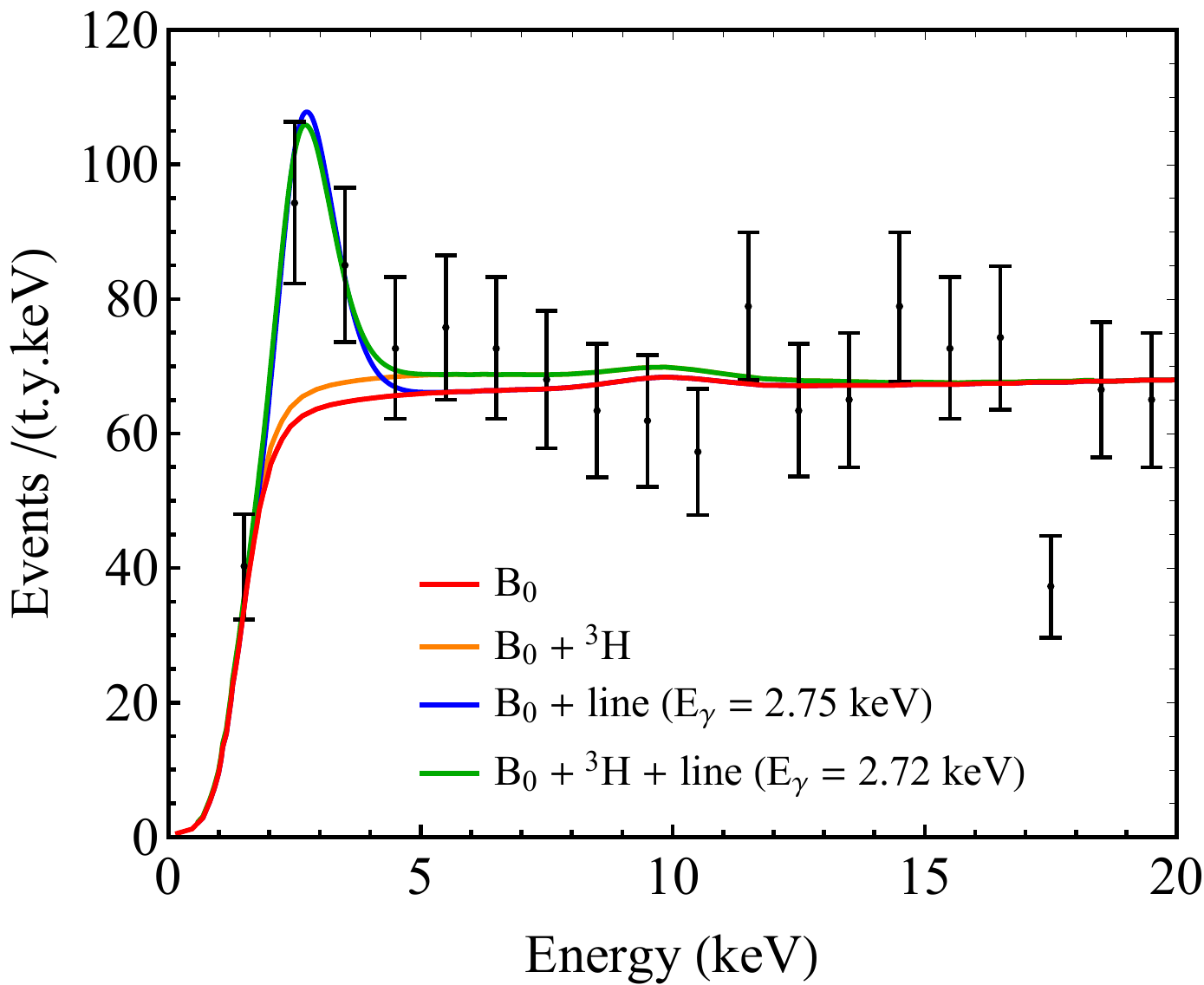}
\caption{The best fit line signal model with (green) and without (blue) the inclusion of a tritium component, compared with the background only event rate (red) and the background plus tritium event rate (orange).}
\label{fig:lineFit}
\end{figure}

\begin{figure}[tbp]
\includegraphics[width=8.cm]{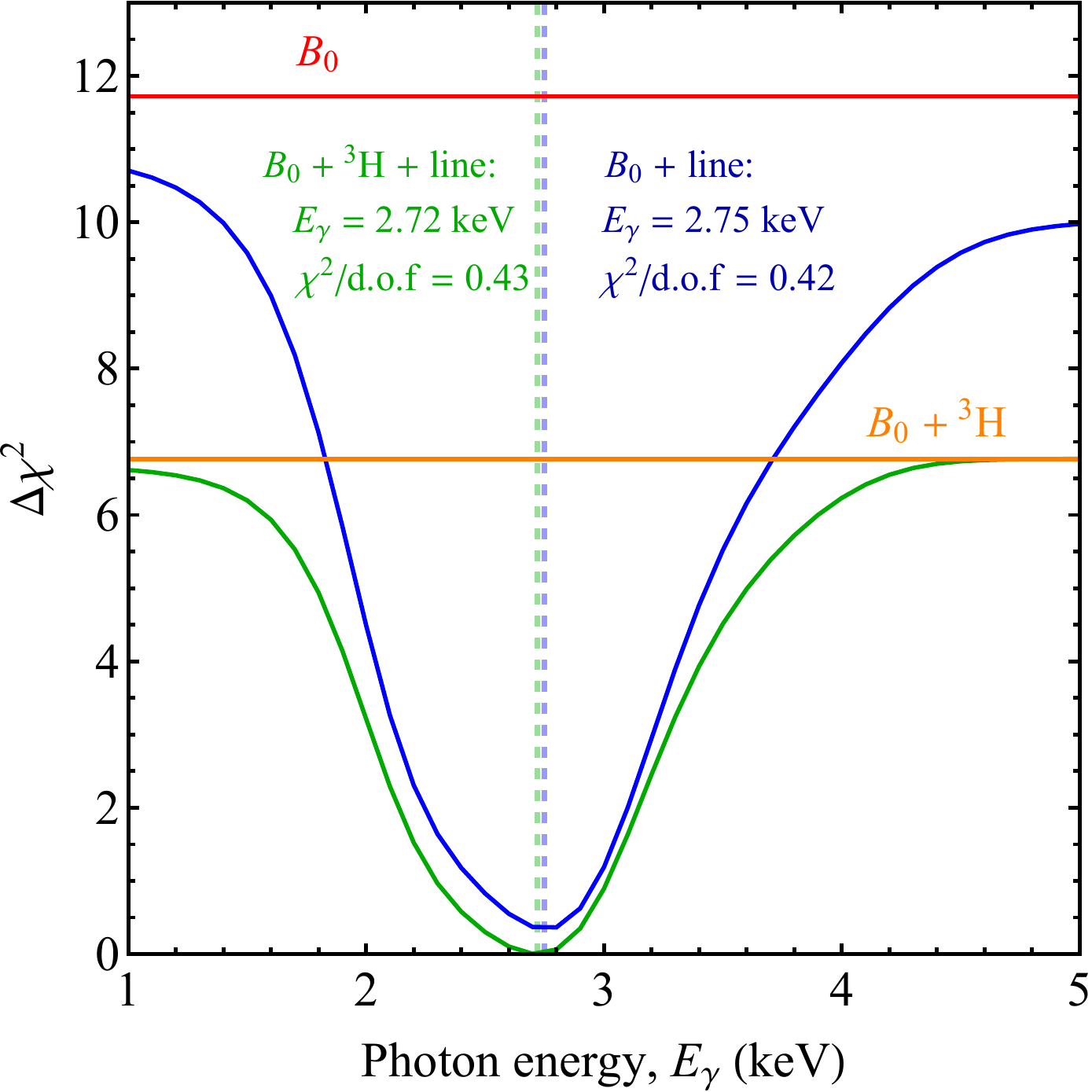}
\caption{The $\Delta\chi^2$ as a function of the photon energy computed based on the 14 lowest-energy bins for luminous dark matter with (green) and without (blue) the inclusion a tritium component. For comparison the background only $\Delta\chi^2$ is also given with (orange) and without (red) the tritium component.}
\label{fig:GoF}
\end{figure}

 The excess events observed by XENON1T are tightly restricted to the energy range of 2-7 keV, with the most significant deviations within just 2 bins from 2-4 keV. Such a narrowly peaked signal can be fit with a mono-energetic photon once smearing due to the detector resolution has been taken into account. The energy resolution of the XENON1T detector can be modeled as a Gaussian with width:
\be
\frac{\sigma(E)}{\mathrm{keV}} = 0.31\sqrt{\frac{E}{\mathrm{keV}}} + 0.0035 \frac{E}{\mathrm{keV}},
\ee
which gives a width of $\sim 18.5\%$ at $E=2.8$ keV, in good agreement with the calibration data \cite{Aprile:2020tmw,arCalibTalk}. The detection efficiency of low energy electron recoils is taken from~\cite{Aprile:2020tmw}. The signal model is defined by two parameters: the line position in energy and the integrated rate. We perform a two parameter fit to the first 14 bins to find the best-fit signal model by minimizing the $\chi^2$ between the data and the signal plus background events. Including additional bins does not affect the best fit point but does help evaluate the relative goodness of fit of the signal models. We find that a line energy of $E_\gamma = 2.75$ keV and rate of $69.8$ events/(tonne$\times$year) provides a good fit to the data: $\chi^2/d.o.f = 0.42$, with $\Delta\chi^2 = 11.4$ compared to the background only model as demonstrated in Fig.~\ref{fig:lineFit}. We also have included the best fit to the excess with LDM plus an unconstrained tritium component. The best fit line signal shifts down slightly to 2.72 keV with a rate of $62.2$ events/(tonne$\times$year), with the tritium mostly contributing to improving the fit in the bins from 4-7 keV. With the tritium inclusion the fit becomes $\chi^2/d.o.f = 0.43$, with $\Delta\chi^2 = 11.7$ compared to the background only rate. Thus a prominent line feature is still a significant component of the excess even when an unconstrained tritium component is added. 
Note that the best-fit energy we obtain is higher than the best-fit obtained by XENON1T's bosonic dark matter fit ($E = 2.3$ keV).  This is due to the bin width of the provided data.   
But the best fit point obtained by the XENON1T collaboration only differs from the best fit point of this analysis by 
half of a bin-width.  In our binned analysis, $\delta = 2.3$ keV is only disfavored by $\Delta \chi^2 \sim 1$, 
so our analysis is consistent with that of XENON1T.  Nothing substantial in the model changes if one chooses 
$\delta = 2.3$ keV, and we see from Figure~\ref{fig:GoF} that this point is strongly preferred to background.

\par For comparison with the best-fit point, we evaluate the $\Delta\chi^2$ for line energies in the range $E_\gamma = 1 - 5$ keV, where the $\Delta\chi^2$ is minimized by finding the best-fit rate and tritium contribution at each energy. The results of this scan are shown in Fig.~\ref{fig:GoF}, we find that line signals in the entire range provide a better fit to the data than the background only model.  Before attempting to explain this excess in terms of BSM physics we stress that the interpretation of the excess as a mono-energetic line stand on their own and could have a SM origin. For example, this line is very close to the x-ray line produced when $^{37}$Ar decays via K-shell electron capture to $^{37}$Cl, which can then relax to its ground state by emitting a 2.8 keV photon~\cite{arCalibTalk}. With a half-life of 35 days, $^{37}$Ar would need to be continuously introduced throughout the data taking period, as no time dependence of the rate was found~\cite{Aprile:2020tmw}. Without a steady source of $^{37}$Ar, this explanation of the excess is strongly disfavored.

\par In the context of LDM, a photon line can provide a viable explanation of the excess so long as a mass splitting of $\sim 2-3~\kev$ is kinematically accessible ($m_\chi \gtrsim 1$ GeV), the multi-scatter veto is evaded and constraints from previous low-threshold analyses are not violated. This includes the XENON1T ionization only (S2-only) analysis~\cite{Aprile:2019xxb}, which constrains both nuclear and electronic recoils, and the standard S1-S2 analysis which constrained nuclear recoils~\cite{Aprile:2018dbl}. The nuclear recoil rate for LDM upscatter is given by,
\be
\frac{dR}{dE_R} = \frac{\rho_\chi}{2m_\chi\mu_{\chi N}}\sigma_{\mathrm{SI}} A^2 F^2(E_R) \int_{v>v_{\mathrm{min}}}\frac{f(v)}{v} dv,
\ee
where $\rho_\chi = 0.3$ GeV/cm$^3$ is the local dark matter density, $\mu_{\chi N}$ is the DM-nucleon reduced mass, $\sigma_{\mathrm{SI}}$ is the spin-independent LDM-nucleon cross section, $A$ is the atomic number of the target (we are assuming identical couplings to neutrons and protons), $F^2(E_R)$ is the nuclear form factor (taken to be of the Helm form~\cite{Lewin:1995rx}) and $f(v)$ is the velocity distribution (taken to be Maxwellian, with a velocity dispersion of $v_0 = 220$ km/s and cutoff at $v_{\mathrm{esc}}=544$ km/s). The kinematics of inelastic scattering require that the incoming DM particle have a minimum velocity given by, 
\be
v_{\mathrm{min}} = \frac{E_R m_T + \delta \mu_{\chi T}}{\sqrt{2 E_R m_T} \mu_{\chi T}}
\ee
where $m_T$ is the mass of the target nucleus and $\mu_{\chi T}$ the reduced mass of the $\chi$ and target nucleus. 

\par
\begin{figure}[tbp]
\includegraphics[width=8.cm]{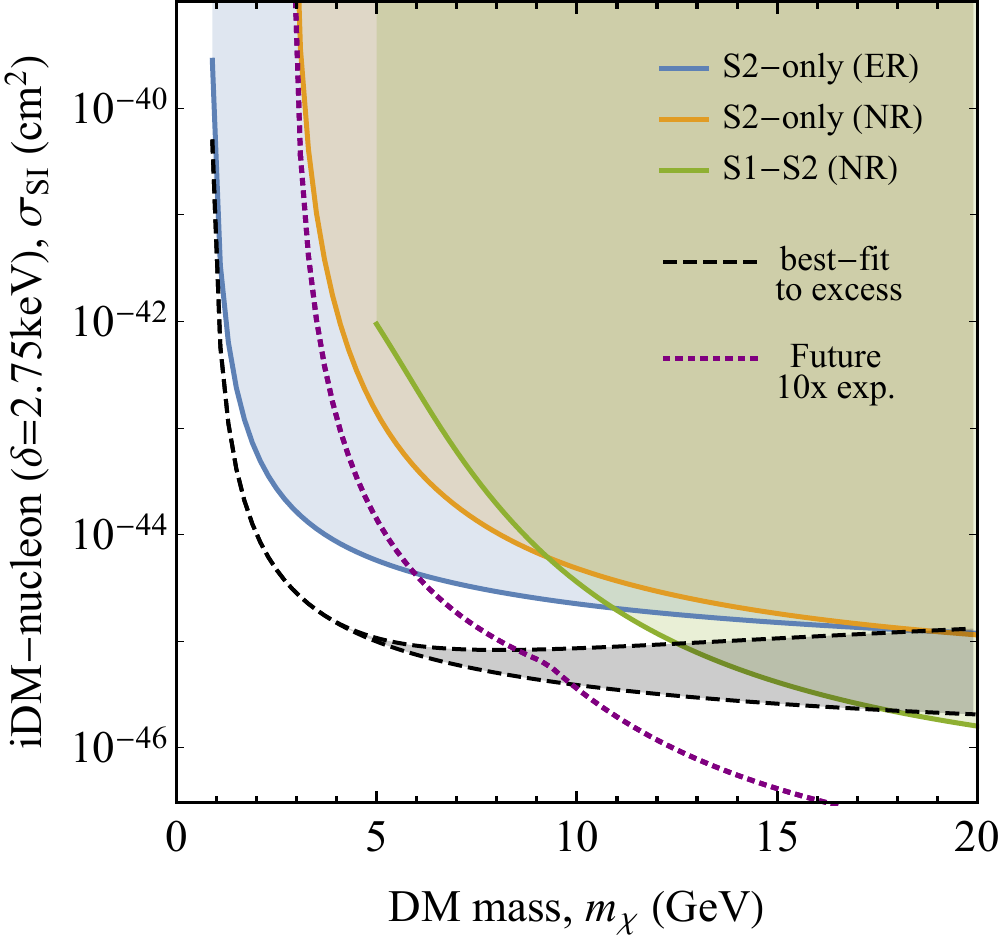}
\caption{Constraints on the spin-independent iDM-nucleon cross section for $\delta =2.75$~keV, derived from other XENON1T analyses~\cite{Aprile:2019jmx,Aprile:2019xxb}. Constraints from S2-only data are given for both electron (blue) and nuclear recoils (orange), while the the S1-S2 results are only used to constrain nuclear recoils (green). The dashed black denotes the approximate cross section required to explain the excess with a 2.75 keV mass splitting. The purple dotted curve denotes the future limits that could be placed directly on the NR signal with 10 times the exposure.}
\label{fig:xe1tconstraints}
\end{figure}

 To check for consistency with previous XENON1T data we perform a single bin analysis where the total upscattering rate is required to be below the total number of observed events in the signal regions of the two NR analyses~\cite{Aprile:2019jmx,Aprile:2019xxb}. For simplicity we perform this analysis for the scenario with LDM only and no tritium. Additionally, the lack of observation of the 
$\sim 2-3~\kev$ line in the S2-only ER data also constrains the cross section. To compute the upper limit we require that the total number of events in the upper two bins of the S2-only analysis in a 22 tonne-day exposure (19 events). These upper bounds are displayed in Fig.~\ref{fig:xe1tconstraints} along with the cross section required to explain the excess (i.e. producing a total rate of 69.8 events/ty). This cross section is given as a range, where the upper limit of the range assumes a worst-case loss of efficiency due to the multi-scatter veto. Note that the loss of efficiency will also affect the constraints we have calculated. A full accounting of the effect of the multi-scatter veto will require a detailed detector simulation, which is beyond the scope of the present work.
\\
\par We find that the LDM scenario is viable for a wide range of DM masses from $\sim$15-17 GeV, down to the kinematic cutoff of 1 GeV where the required cross section is on the cusp of the S2-only ER bound. In the near future XENONnT~\cite{Brown-XENONnT, Evan-XENON1T} will begin operating. With three times the fiducial mass of XENON1T, it will be able to collect 10 times the exposure of XENON1T in a few years. Assuming a commensurate reduction in the background rate, such an exposure will directly probe the nuclear recoils of this LDM scenario down to 10 GeV in dark matter mass.

\section{Future Prospects} \label{sec:prospects}

We briefly discuss some avenues for probing this scenario 
with data from future experiments.

\begin{itemize}
    \item{{\it Direct detection spectrum}: 
    Upcoming direct detection 
    experiments should be able to distinguish between this 
    scenario and other BSM scenarios, and possible tritium 
    backgrounds.  In particular, as energy resolution 
    improves, the peak arising from LDM will 
 become increasingly sharp, and therefore more easily distinguishable from other possibilities. }
    \item{{\it Multi-channel direct detection signal}:
    Interestingly, because $m_\chi$ can be as large as 
    $15~\gev$, future xenon-based direct detection 
    experiments could potentially see a nuclear recoil 
    signal.  
    If the tail of the nuclear recoil spectrum from inelastic scattering is above threshold, future experiments such as LZ~\cite{Akerib:2019fml}, XENONnT~\cite{Brown-XENONnT, Evan-XENON1T} or PandaX-4T~\cite{Zhang:2018xdp} may see both nuclear recoils and the decay signal, which would be a powerful cross check.  In particular, this signal could also 
    help distinguish this scenario from that of bosonic 
    dark matter absorption, which also yields a monoenergetic 
    peak.
    Specific model details would naturally arise when considering future signals. For example, some models produce lifetimes that would preclude observation of prompt decays within the detector, so one would see 
    nuclear recoils and decay photons, but they would not be from the same event (see, for example, \cite{Baryakhtar:2020rwy} for a recent discussion).}
    \item{{\it Diurnal modulation}:  
    If the decay length is of 
   order the length of the overburden (${\cal O}(10^3)~\m$) or greater, then more events will be observed when the dark 
   matter wind passes through the Earth (yielding a larger 
   volume for scattering), while fewer events will be 
   observed if the dark matter wind comes from above 
   the detector.  This diurnal modulation was discussed 
   in the context of LDM in~\cite{Eby:2019mgs}.
}
    \item{{\it Collider production}:  
    $\chi$ or $\chi'$ can be produced at beam experiments, yielding either photon or missing 
    energy signatures.  
    For example, the $pp \rightarrow \chi \chi' j$ process 
    will yield a monojet and missing energy 
    signal~\cite{Aaboud:2017phn, Khachatryan:2014rra} if the 
    $\chi'$ lifetime is sufficiently large. 
    For a short lifetime (decay within the detector), it will produce monophoton final state~\cite{Baek:2018hnc,Sirunyan:2017ewk}. However, in order to be observed at the LHC, the boosting of the $\chi'$ would have to be substantial, resulting in a $\sigma_{\chi p}$ cross-section required by the fit that would be suppressed and mostly out of reach for the LHC.
    }
    \item{{\it Beam-dump/fixed-target experiments}: $\chi'$ can be produced in beam-dump/fixed target experiments and if it is long-lived then an energetic photon spectrum 
    could be seen at FASER~\cite{Feng:2017uoz, Ariga:2018zuc, Ariga:2018pin, Ariga:2018uku}, SHiP~\cite{Anelli:2015pba, Alekhin:2015byh}, SeaQuest~\cite{Aidala:2017ofy, Berlin:2018pwi}, or other 
    displaced detectors.  The production cross section, however, will be dependent on the model dependent details of the interaction 
    between dark matter and the Standard Model.}

\end{itemize}


\section{Summary} \label{sec:summary} 

XENON1T has recently reported an interesting unexplained 
excess of 
electron recoil events, with typical energies of a 
few keV.  Although this excess can potentially be 
explained by a tritium background, there has naturally 
been interest in BSM explanations of this signal.  We have
shown that this signal can be produced by a species of 
Luminous Dark Matter with mass in the $\sim 1-15~\gev$ range, 
with a mass splitting between the heavy and light states of 
$2.75~\kev$.  If the dark matter scatters inelastically 
with nuclei in the detector or the surrounding rock, then 
the heavier state can decay back to the light state within 
the detector, emitting one or more photons with an energy 
of $\sim 2.75~\kev$.  Including the effects of the energy 
resolution, this model is a good fit to the data: $\chi^2/d.o.f = 0.42$ with $\Delta\chi^2 = 11.4$ compared to the background only case.  When we include a tritium contribution, the best fit line value shifts down to 2.72 keV, and the fit becomes: $\chi^2/d.o.f. = 0.43$ with $\Delta\chi^2 = 11.7$ compared to the background only model. This is 
a very general framework; the fit to the data depends primarily 
on the mass splitting and the decay channel (to photons), 
but has very little specific model dependence on the microphysics. 

This scenario can be probed with future data from direct 
detection experiments, which can be used to distinguish 
this scenario from other BSM scenarios, as well as from 
the tritium background. As an example, the LDM structure allows for a possible distinctive diurnal feature, or multi-channel detection, etc., which can be searched for in future experiments. In addition, collider, beam-dump/fixed target experiments can also provide interesting 
signals, but these are much more dependent on the details 
of dark matter interactions with the Standard Model.

{\bf \emph{Acknowledgements}} We are grateful for helpful discussion with Rafael Lang and Cara Giovanetti.
NFB and JLN were supported in part by the Australian Research Council. JBD acknowledges support from the National Science Foundation under Grant No. NSF PHY-1820801. BD and SG acknowledge support from  the DOE Grant No. DE-SC0010813. The work of JK is supported in part by DOE grant DE-SC0010504.

\bibliography{XenonMigdal}

\end{document}